\documentclass[epj]{svjour}
\usepackage{amsmath, amssymb, fullpage, graphics, color}

\newcommand{\bA}{\mathbf{A}}
\newcommand{\bB}{\mathbf{B}}
\newcommand{\bC}{\mathbf{C}}
\newcommand{\bF}{\mathbf{F}}
\newcommand{\bU}{\mathbf{U}}
\newcommand{\bu}{\mathbf{u}}

\newcommand{\bbf}{\mathbf{f}}

\begin{document}
\title{Hyperbolic model of internal solitary waves in a three-layer stratified fluid}
\author{Alexander Chesnokov\inst{1,2} \and Valery Liapidevskii\inst{1,2}
}                     
%
%
\institute{Lavrentyev Institute of Hydrodynamics SB RAS, 15 Lavrentyev Ave., Novosibirsk 630090, Russia \and Novosibirsk State University, 1 Pirogova Str., Novosibirsk 630090, Russia}
\mail {chesnokov@hydro.nsc.ru}
\date{Received: date / Revised version: date}
%
\abstract{
We derive a new hyperbolic model describing the propagation of internal waves in a stratified shallow water with a non-hydrostatic pressure distribution. The construction of the hyperbolic model is based on the use of additional `instantaneous' variables. This allows one to reduce the dispersive multi-layer Green--Naghdi model to a first-order system of evolution equations. The main attention is paid to the study of three-layer flows over uneven bottom in the Boussinesq approximation with the additional assumption of hydrostatic pressure in the intermediate layer. The hyperbolicity conditions of the obtained equations of three-layer flows are formulated and solutions in the class of travelling waves are studied. Based on the proposed hyperbolic and dispersive models, numerical calculations of the generation and propagation of internal solitary waves are carried out and their comparison with experimental data is given. Within the framework of the proposed three-layer hyperbolic model, a numerical study of the propagation and interaction of symmetric and non-symmetric soliton-like waves is performed.
} 
\maketitle
\section{Introduction}\label{sec:1}
Internal solitary waves (ISWs) that propagate along a density interface can be discovered at many locations in the stratified oceans \cite{Helfrich_2006,Jackson_2004}. These waves are important as their energy and mass transport can produce a substantial impact on any offshore structures, marine biology and geology. The diversity of ISWs shapes is determined by the stratification of water, the bottom profile, and various mechanisms of the non-linear wave generation. The observed ISWs are most frequently mode-1 waves that displace isopycnals in one direction and can either be waves of elevation or, more typically, waves of depression. Mode-2 ISWs displace isopycnals in opposite directions. The last decade has seen a growth of observations of mode-2 ISWs which have isopycnals that expand away and contract towards the pycnocline centre \cite{Shroyer_2010,Ramp_2012,Silva_2015}. Large amplitude mode-2 ISWs have unique properties, in particular regions of internal recirculation that enable mass transport over large distances. Therefore, a comprehensive study of the generation, propagation and transformation of mode-2 ISWs is of considerable interest.

In addition to the field observations of ISWs mentioned above, significant progress has been made in describing this phenomenon through theoretical analysis \cite{Akylas_1992,Choi_2000,Antuono_2009}, numerical simulations \cite{Salloum_2012,Zhang_2018}, and laboratory experiments \cite{Brandt_2014,Carr_2015}. It should be noted that many studies of ISWs are based on combined methods and approaches. In \cite{Yuan_2018} the propagation of mode-2 ISWs over a slope-shelf topography was investigated using both analysis and numerical simulations. The existence of a long mode-1 wave ahead of mode-2 ISWs was observed and it was found that this process cannot be described by the Korteweg--de Vries theory. The dynamics and energetics of a head-on collision of ISWs with trapped cores propagating in a thin pycnocline were studied numerically in \cite{Maderich_2017} within the framework of the Navier--Stokes equations for a stratified fluid. The peculiarity of this collision is that it involves trapped masses of a fluid. In recent works \cite{Deepwell_2017,Deepwell_2019}, the main features of the interaction of mode-2 ISWs with a narrow and broad isolated topography were investigated applying both numerical simulations and laboratory experiments. It was found that the mode-2 incident wave generates multiple internal waves and wave types from the interaction with a broad ridge at sufficiently high amplitude and wave speed. The decaying mode-2 ISWs over a bottom step in a computational tank filled with a three-layer stratified fluid was established in \cite{Terletska_2016}. For numerical simulation of the ISWs evolution, the 2D or 3D non-linear Navier--Stokes equations in the Boussinesq approximation are often used. This allows one to obtain detailed information about the stratified fluid flow, but requires the use of high-performance computing cluster for calculations. 

An important role in the study of non-linear surface and internal waves in a stratified fluid is played by multi-layer models of the second-order approximation of shallow water theory. For small aspect ratio of the thickness of each fluid layer to typical wavelength, a strongly non-linear 1D model to describe the evolution of finite amplitude long internal waves in a multi-layer system was proposed in \cite{Choi_2000}. Various modifications of this non-linear dispersive model, mainly for a three-layer fluid, were considered in \cite{Gavrilov_2012,Liapidevskii_2017,Liapidevskii_2018,Kukarin_2019}. In these works, theoretical results were compared with experimental data and field observations of the propagation of internal waves in the coastal zone. It was shown that dispersion models of three-layer shallow water made it possible to correctly describe the main features of the transformation of large-amplitude internal waves. In particular, non-symmetric mode-2 ISWs were obtained and verified by comparison with the laboratory experiment \cite{Liapidevskii_2018}. Recently, a strongly non-linear long-wave model for large amplitude internal waves in a three-layer flow between two rigid boundaries was considered in \cite{Barros_2020}. Emphasis was given to the solitary waves of the second baroclinic mode and their strongly non-linear characteristics that fail to be captured by weakly non-linear models.

One of the major numerical challenges in solving dispersive shallow water equations consist in the resolution of an elliptic problem at each time instant and realization of non-reflecting conditions at the boundary of the calculation domain \cite{LeM_G_H_2010}. An alternative formulation of dispersive models proposed in \cite{LT00,Antuono_2009}  within the framework of hyperbolic equations allow one to avoid these problems. In \cite{Liapidevskii_2008,Favrie_Gavr_2017,Chesn_Ng_2019}, this approach was further developed and applied to simulate the evolution of non-linear surface waves over uneven topography. The main goal of this paper is to obtain and study the hyperbolic approximation of non-linear shallow water equations for three-layer non-hydrostatic flows and apply this model to describe ISWs. As far as the authors know, this approach to the derivation of hyperbolic equations for describing stratified non-hydrostatic flows is used for the first time.

The remainder of this paper is organized as follows. In Section~\ref{sec:2}, we recall a non-linear model describing three-layer stratified shallow-water flows with a non-hydrostatic pressure distribution in the outer layers. Then, using additional `instantaneous' variables, we derive a hyperbolic system that approximates the considered three-layer dispersive model. In Section~\ref{sec:3}, we study the solutions of the hyperbolic model in the form of travelling waves and formulate the necessary conditions for the existence of a ISW. We present a solution of the model describing a non-symmetric ISW of mode-2. This solution is verified by comparison with the known experimental data. In Section~\ref{sec:4}, we present the results of numerical simulation of non-stationary ISWs in a three-layer fluid. At first, we demonstrate that the numerical solution of the hyperbolic model approximates with high accuracy the solution of the original dispersive model. Then, in the framework of hyperbolic equations, we simulate the transformation of mode-2 ISWs over a broad isolated ridge and the interaction of two mode-2 symmetric and non-symmetric ISWs. Finally, we draw some conclusions.

\section{Governing equations}\label{sec:2}

We consider strongly non-linear internal gravity waves propagating in a stratified fluid. A mathematical model describing the time evolution of large amplitude internal waves in a multilayer stratified fluid was proposed in \cite{Choi_2000}. This model follows from the Euler equations under the sole assumption that the waves are long compared to the undisturbed thickness of the fluid layers. In this work we make some additional assumptions. We restrict our consideration to three-layer flows over a mild slope bottom topography. The lower and upper homogeneous fluid layers with densities $\rho^-$ and $\rho^+$ are separated by an interlayer with the density $\bar{\rho}$. In what follows, we assume that $\rho^+<\bar{\rho}<\rho^-$, and the pressure in the interlayer obeys the hydrostatic law. This means that the thickness of the interlayer is small compared to the thicknesses of the lower and upper non-hydrostatic layers. The introduction of such interlayer is reasonable due to the generation of small-scale motions at the interface of homogeneous layers during the passage of internal waves of finite amplitude~\cite{LT00}. 

Under these assumptions in the Boussinesq approximation the governing equations can be written as~\cite{Liapidevskii_2017}
\begin{equation} \label{eq:3L}
\begin{array}{l} \displaystyle
h_t+(uh)_x=0, \quad \eta_t+(v\eta)_x=0, \quad \zeta_t+(w\zeta)_x=0, \\[3mm]\displaystyle
u_t+uu_x+bh_x+\bar{b}\eta_x+p_x+ \frac{\varepsilon^2}{3h}\Big(h^2\frac{d_-^2h}{dt^2}\Big)_x =-bZ_x, \\[3mm]\displaystyle
v_t+vv_x+\bar{b}h_x+\bar{b}\eta_x+p_x=-\bar{b}Z_x, \quad
w_t+ww_x+p_x+ \frac{\varepsilon^2}{3\zeta}\Big(\zeta^2\frac{d_+^2\zeta}{dt^2}\Big)_x=0.
\end{array}
\end{equation}
Here $h$, $\eta$ and $\zeta$ are the thickness of the lower, intermediate, and upper layers; $u$, $v$, and $w$ are the velocities in these layers; $p\rho^+$ is the pressure at the upper boundary of the flow region; $\varepsilon\ll 1$ is the dimensionless long wave parameter; $z=Z(x)$ is the bottom topography; $d_\mp/dt$ stand for material derivatives
\[ \frac{d_-}{dt}=\frac{\partial}{\partial t}+u\frac{\partial}{\partial x}, \quad 
\frac{d_+}{dt}=\frac{\partial}{\partial t}+w\frac{\partial}{\partial x}\,. \]
The constants $b$ and $\bar{b}$ determine the buoyancy coefficients as follows
\[ b=g(\rho^--\rho^+)/\rho^+, \quad \bar{b}=g(\bar{\rho}-\rho^+)/\rho^+, \]
where $g$ is the gravity acceleration. 

As it was mentioned above, we apply here a mild slope approximation. This means that the dimensionless bottom variation is weak \cite{Serre_1953,G_L_Ch_2019}: $z=Z(\varepsilon^\gamma x)$, $\gamma>0$. Due to this fact the terms $\varepsilon^2 Z_x$ and $\varepsilon^2 Z_{xx}$ can be neglected during the derivation of system~(\ref{eq:3L}). 

It should be noted that in the Boussinesq approximation the upper boundary is fixed
\begin{equation} \label{eq:H} 
h+\eta+\zeta+Z(x)=H+Z(x)=H_0 \equiv{\rm const}.
\end{equation}
This allows us to obtain one more integral of (\ref{eq:3L}) and exclude two equations from this system. It follows from (\ref{eq:H}) and the first three equations (\ref{eq:3L}) that the flow rate is a function of the variable $t$
\begin{equation} \label{eq:Q}  
uh+v\eta+w\zeta=Q(t). 
\end{equation}
Formulas (\ref{eq:H}) and (\ref{eq:Q}) determine the flow parameters in the interlayer as follows
\begin{equation} \label{eq:eta_v} 
\eta=H-h-\zeta, \quad v=\frac{Q-uh-w\zeta}{\eta}\,. 
\end{equation}

It is easy to see that using the first and third equations in (\ref{eq:3L}), we can rewrite the dispersive terms in the form
\begin{equation} \label{eq:disp-corr} 
\begin{array}{l} \displaystyle
\frac{1}{h}\frac{\partial }{\partial x}\Big(h^2\frac{d_-^2h}{dt^2}\Big)= \frac{\partial }{\partial x} \Big(h\frac{d_-^2h}{dt^2}+ \frac{1}{2}\Big(\frac{d_-h}{dt}\Big)^2\Big) +L_-, \\[4mm]\displaystyle 
\frac{1}{\zeta}\frac{\partial }{\partial x}\Big(\zeta^2\frac{d_+^2\zeta}{dt^2}\Big)= 
\frac{\partial }{\partial x} \Big(\zeta\frac{d_+^2\zeta}{dt^2}+\frac{1}{2}\Big(\frac{d_+\zeta}{dt}\Big)^2\Big) +L_+,
\end{array} 
\end{equation}
where
\[ L_-=(h_t u_{xx}-h_x u_{xt})h, \quad L_+=(\zeta_t w_{xx}-\zeta_x w_{xt})\zeta. \]
In view of (\ref{eq:disp-corr}), equations (\ref{eq:3L}) admit a divergent representation up to the terms $L_\mp$. Let us note that the functions $L_\mp$ vanish in the class of stationary solutions for flows over uneven bottom and travelling waves for flows over flat topography. In the general case, substituting representation~(\ref{eq:disp-corr}) into equations~(\ref{eq:3L}) and dropping the terms $L_\mp$ lead to an error of the order $\varepsilon^2$ or less. Such modification of system~(\ref{eq:3L}) was proposed in \cite{Gavrilov_2013}. 

In this case the governing equations take the conservative form
\begin{equation} \label{eq:3L-mod}
\begin{array}{l} \displaystyle
h_t+(uh)_x=0, \quad \zeta_t+(w\zeta)_x=0, \\[3mm]\displaystyle
(u-v)_t+\Big(\frac{u^2-v^2}{2}+(b-\bar{b})h+ \frac{\varepsilon^2 h}{3}\frac{d_-^2h}{dt^2}+ \frac{\varepsilon^2}{6}\Big(\frac{d_-h}{dt}\Big)^2 \Big)_x =-(b-\bar{b})Z_x, \\[3mm]\displaystyle
(w-v)_t+\Big(\frac{w^2-v^2}{2}+\bar{b}\zeta+ \frac{\varepsilon^2 \zeta}{3}\frac{d_+^2\zeta}{dt^2}+\frac{\varepsilon^2}{6}\Big(\frac{d_+\zeta}{dt}\Big)^2 \Big)_x=0
\end{array}
\end{equation}
with closing relations (\ref{eq:eta_v}). 

Both models, original (\ref{eq:3L}) and modified (\ref{eq:3L-mod}), admit a uniform presentation convenient for a numerical treatment of non-stationary problems \cite{Liapidevskii_2017,Gavrilov_2013}:
\begin{equation} \label{eq:3L-num}
\begin{array}{l} \displaystyle
h_t+(uh)_x=0, \quad \zeta_t+(w\zeta)_x=0, \\[3mm]\displaystyle
K_t+\Big(Ku-\frac{(u-v)^2}{2}+(b-\bar{b})h-
\frac{\varepsilon^2}{2(2n+1)}h^2u_x^2 \Big)_x=-(b-\bar{b})Z_x, \\[3mm]\displaystyle
R_t+\Big(Rw-\frac{(w-v)^2}{2}+\bar{b}\zeta-
\frac{\varepsilon^2}{2(2n+1)}\zeta^2w_x^2 \Big)_x=0.
\end{array}
\end{equation}
Here
\begin{equation} \label{eq:K_R} 
K=u-v-\frac{\varepsilon^2}{3h^{1-n}}\big(h^{3-n} u_x\big)_x, \quad 
R=w-v-\frac{\varepsilon^2}{3\zeta^{1-n}}\big(\zeta^{3-n} w_x\big)_x.
\end{equation}
We take $n=0$ for model (\ref{eq:3L}) and $n=1$ for modified equations (\ref{eq:3L-mod}). The flow parameters in the interlayer are determined by formulas~(\ref{eq:eta_v}). It should be noted that for all numerical tests considered below, there is no visible difference in the results obtained on the basis of system~(\ref{eq:3L-num})--(\ref{eq:K_R}) with $n=0$ and $n=1$.

Numerical solution of equations~(\ref{eq:3L-num})--(\ref{eq:K_R}) can be divided into two successive steps: i) time evolution of the conservative variables $(h, \zeta, K, R)$ using the Godunov-type method for system~(\ref{eq:3L-num}); ii) resolution of the second order ODEs (\ref{eq:K_R}) to find the velocities $u$ and $w$. This approach was proposed in \cite{LeM_G_H_2010} for numerical solution of the Green--Naghdi equations and was used, for instance, in \cite{Liapidevskii_2017,G_L_Ch_2019} for modelling stratified shallow water flows with a non-hydrostatic pressure distribution. 

As it was mentioned above, the major numerical challenges in solving dispersive shallow water equations consist in the resolution of an elliptic problem at each time instant and realization of non-reflecting boundary conditions. Below we present and study a hyperbolic approximation of the modified dispersive model (\ref{eq:3L-mod}), which allow one to avoid these problems.

\subsection{Hyperbolic approximation}\label{subsec:2.1}

Following \cite{Liapidevskii_2008,Chesn_Ng_2019}, we introduce new instantaneous variables $\tilde{h}$, $\tilde{u}$, $\tilde{\zeta}$ and $\tilde{w}$ so that 
\begin{equation} \label{eq:inst_var} 
\frac{d_-\tilde{h}}{dt}=\tilde{u}, \quad 
\frac{d_-\tilde{u}}{dt}=\frac{\alpha \bar{b}}{h}(h-\tilde{h}),
\quad \frac{d_+\tilde{\zeta}}{dt}=\tilde{w}, \quad \frac{d_+\tilde{w}}{dt}= 
\frac{\alpha \bar{b}}{\zeta}(\zeta-\tilde{\zeta}). 
\end{equation}    
Here $\alpha>0$ is the non-dimensional parameter. Then we replace the dispersive terms in equations~(\ref{eq:3L-mod}) as follows
\begin{equation} \label{eq:appr_deriv}  
\frac{d_-h}{dt}\to\tilde{u}, \quad 
\frac{d_-^2h}{dt^2}\to\frac{\alpha \bar{b}}{h}(h-\tilde{h}), \quad 
\frac{d_+\zeta}{dt}\to\tilde{w}, \quad 
\frac{d_+^2\zeta}{dt^2}\to\frac{\alpha \bar{b}}{\zeta}(\zeta-\tilde{\zeta}).
\end{equation}   
As a result, we obtain the following first-order system of balance laws 
\begin{equation} \label{eq:3L-hyp}
\begin{array}{l} \displaystyle
h_t+(uh)_x=0, \quad s_t+\Big(\frac{u^2-v^2}{2}+a_1 h-
\frac{\alpha\varepsilon^2}{3}\bar{b}\tilde{h}+\frac{\varepsilon^2}{6}\tilde{u}^2 \Big)_x =-(b-\bar{b})Z_x, \\[3mm]\displaystyle
\zeta_t+(w\zeta)_x=0, \quad r_t+\Big(\frac{w^2-v^2}{2}+a_2\zeta-
\frac{\alpha\varepsilon^2}{3}\bar{b}\tilde{\zeta}+    
\frac{\varepsilon^2}{6}\tilde{w}^2 \Big)_x =0, \\[3mm]\displaystyle
(h\tilde{h})_t+(uh\tilde{h})_x=\tilde{u}h, \quad 
(\tilde{u}h)_t+(u\tilde{u}h)_x=\alpha \bar{b}(h-\tilde{h}), \\[3mm]\displaystyle
(\zeta\tilde{\zeta})_t+(w\zeta\tilde{\zeta})_x=\tilde{w}\zeta, \quad 
(\tilde{w}\zeta)_t+(w\tilde{w}\zeta)_x=\alpha \bar{b}(\zeta-\tilde{\zeta}),
\end{array}
\end{equation}
where $s=u-v$ and $r=w-v$ are the relative velocities in the lower and upper layers, the coefficients $a_i$ are of the form
\[ a_1=\Big(\frac{b}{\bar{b}}-1+\frac{\alpha\varepsilon^2}{3}\Big)\bar{b}, \quad
a_2=\Big(1+\frac{\alpha\varepsilon^2}{3}\Big)\bar{b}\,. \]
Here we present equations for the instantaneous variables in a conservative form (the last four equations in (\ref{eq:3L-hyp})) since it is convenient for numerical treatment. Taking into account formulas (\ref{eq:eta_v}), we can express velocities $u$, $v$ and $w$:
\[ u=\frac{Q+s\eta+(s-r)\zeta}{H}, \quad 
v=\frac{Q-sh-r\zeta}{H}, \quad
w=\frac{Q+r\eta-(s-r)h}{H}. \]

As the parameter $\alpha$ increases, the solutions of equations~(\ref{eq:3L-hyp}) approximate the solutions of dispersive system (\ref{eq:3L-mod}). This follows from the definition of instantaneous variables by formulas~(\ref{eq:inst_var}). The construction of a similar approximation for equations~(\ref{eq:3L}) is also possible. Our choice of system~(\ref{eq:3L-mod}) is due to the fact that it has a divergent form and, consequently, the obtained approximation~(\ref{eq:3L-hyp}) is also written in the form of conservation laws. 

In what follows, we take $\varepsilon=1$ for all proposed models.

\subsection{Characteristics of equations~(\ref{eq:3L-hyp})}\label{subsec:2.2}

Let us find the characteristics of system~(\ref{eq:3L-hyp}) and the conditions for its hyperbolicity. Obviously, there are four contact characteristics $dx/dt=u$ and $dx/dt=w$, each of them has a multiplicity of two. Equations~(\ref{eq:3L-hyp}) can be written in the form
\begin{equation} \label{eq:U-form}  
\bU_t+\bA\bU_x=\bF, 
\end{equation}  
where $\bU=(h, \zeta, s, r, \tilde{h}, \tilde{\zeta}, \tilde{u}, \tilde{w})^{\rm T}$ is the vector of unknown variables, $\bA$ is the $8\times 8$ matrix, and $\bF$ is the right-hand side. The eigenvalues of $\bA(\bU)$ are determined by equation
\[ \chi(\lambda)=(u-\lambda)^2(w-\lambda)^2\hat{\chi}(\lambda)=0, \] 
where
\begin{equation} \label{eq:chi} 
\begin{array}{l} \displaystyle
\hat{\chi}(\lambda)=((u-\lambda)^2-a_1 h)((w-\lambda)^2-a_2\zeta)\eta+  \\[3mm]\displaystyle \quad\quad\quad
+\Big(((u-\lambda)^2-a_1 h\big)\zeta+ ((w-\lambda)^2-a_2\zeta)h\Big)(v-\lambda)^2.
\end{array} 
\end{equation}
System (\ref{eq:3L-hyp}) is hyperbolic if the polynomial $\hat{\chi}(\lambda)$ has four real different roots. 

If the thickness of one of the layers vanishes, the polynomial $\hat{\chi}(\lambda)$ is represented as the product of two polynomials of the second degree. In this case, obtaining the condition for the existence of four real roots of the equation $\hat{\chi}(\lambda)=0$ is not difficult. For example, if the thickness $\eta$ is zero, then the inequality $(u-w)^2<(a_1+a_2)H$ ensures the existence of four real roots of the equation $\hat{\chi}(\lambda)=0$. To fulfil this inequality, it is needed to choose the parameter $\alpha$ as follows
\[ \alpha>\max\Big\{0, \frac{3((u-w)^2-bH)}{2\bar{b}H}\Big\}\,. \]
Further, we consider the general case when the thickness of each layer is positive.

An insightful geometric interpretation of the characteristics proposed in \cite{Ovs_79,Chesn_2017} for two-layer hydrostatic flows can be applied here. We introduce new variables $\Phi$ and $\Psi$ by the formulas
\begin{equation} \label{eq:Phi-Psi} 
\Phi=\frac{u-\lambda}{\sqrt{a_1 h}}, \quad \Psi=\frac{w-\lambda}{\sqrt{a_2 \zeta}}\,.
\end{equation}
Then equation $\hat{\chi}(\lambda)=0$ can be rewritten in the form
\begin{equation} \label{eq:curve}  
F(\Phi,\Psi)\equiv(\Phi^2-1)(\Psi^2-1)+\Big(\frac{\Phi^2-1}{a_2\eta} +\frac{\Psi^2-1}{a_1\eta})\Big) 
{(u-v-\Phi\sqrt{a_1 h})^2}=0. 
\end{equation}
The variables $\Phi$ and $\Psi$ by virtue of (\ref{eq:Phi-Psi}) are related by the expression
\begin{equation} \label{eq:line}  
\Psi=\Phi\sqrt{\frac{a_1 h}{a_2 \eta}}-\frac{u-w}{\sqrt{a_2\zeta}}\,.
\end{equation}
The number of real roots of equation $\hat{\chi}(\lambda)=0$ is determined by the number of intersections of the curve~(\ref{eq:curve}) with the straight line~(\ref{eq:line}). Each point of interaction yields a sonic characteristic with the slope $\lambda=w-\Psi\sqrt{a_2\zeta}$. 

\begin{figure}[t]
	\begin{center}
		\resizebox{.8\textwidth}{!}{\includegraphics{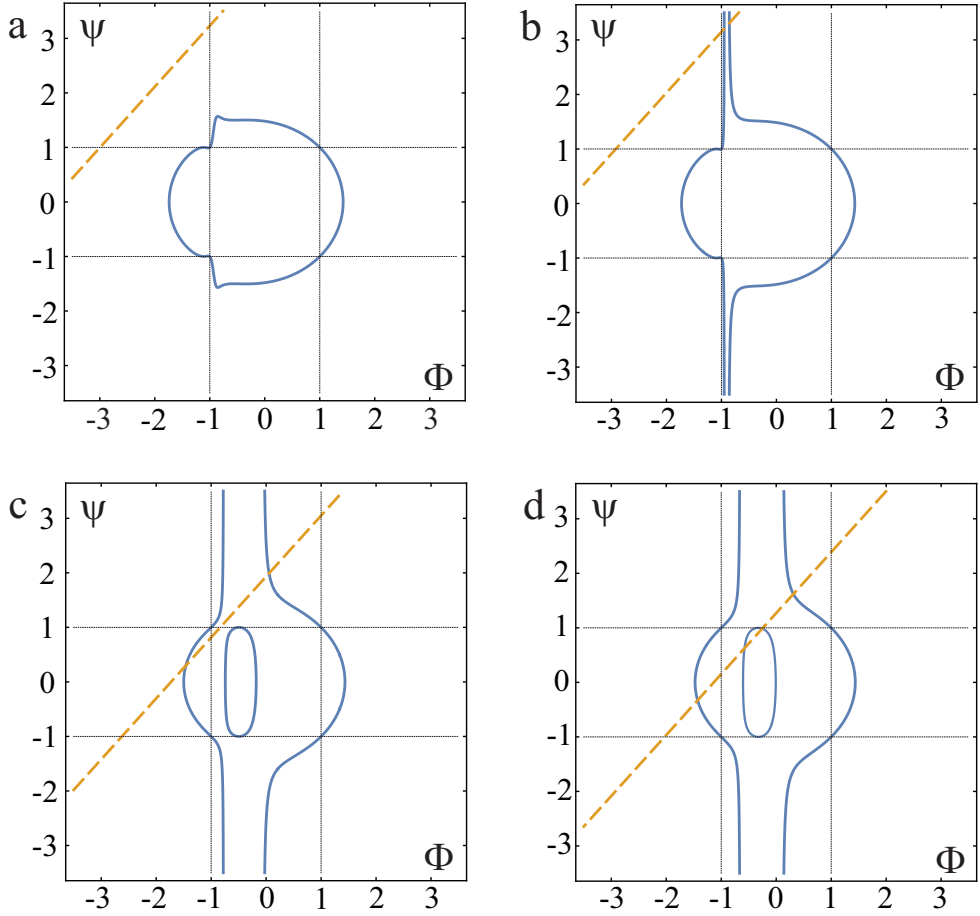}}\\[0pt]
		{\caption{The curves (\ref{eq:curve}) and the straight lines (\ref{eq:line}) in the $(\Phi,\Psi)$--plane for $h=5$, $\eta=1$, $\zeta=4$, $u=-1$, $v=-1/3$, $w=12/9$, $b=0.058$ and $\bar{b}=0.029$. (a) --- $\alpha=4.5$; (b) --- $\alpha=4.8$; (c) --- $\alpha=35$; (d) --- $\alpha=85$.} \label{fig:fig_1}} 
	\end{center}
\end{figure}

In the $(\Phi, \Psi)$-plane, equation~(\ref{eq:curve}) describes a fourth-order curve. Let us study its  properties. The characteristic form of curve (\ref{eq:curve}) for fixed flow parameters and various values of $\alpha$ is shown in Fig.~\ref{fig:fig_1} by solid curves. To plot the graphs, we choose the following parameters of the three-layer flow: $h=5$, $\eta=1$, $\zeta=4$ (m); $u=-1$, $v=-1/3$, $w=12/9$ (m/s); $b=0.058$, $\bar{b}=0.029$ (m/s$^2$). 

If the velocities in the layers are different and the parameter $\alpha$ is small, then equation $\hat{\chi}(\lambda)=0$ may not have real roots (Fig.~\ref{fig:fig_1}\,(a)). We note that for $\Psi^2\to\infty$ the equation $F(\Phi,\Psi)=0$ yields 
\[ \Phi\to\Phi_{1,2}=\frac{u-v}{h+\eta}\sqrt{\frac{h}{a_1}}\mp
\sqrt{\Big(1-\frac{(u-v)^2}{(h+\eta)a_1}\Big)\frac{\eta}{h+\eta}}\,. \]
This means that for $a_1>(u-v)^2/(h+\eta)$ the curve has two vertical asymptotes $\Phi=\Phi_{1,2}$ for $\Psi^2\to\infty$ and, consequently, there are (at least) two points of intersection with the straight line (see Fig.~\ref{fig:fig_1}\,(b)). For $\Psi^2=1$, it follows from equation~(\ref{eq:curve}) that $\Phi^2=1$ and $\Phi^2=\Phi_*^2=(u-v)^2/(a_1 h)$. Therefore, inequality $|\Phi_*|<1$ (or $a_1>(u-v)^2/h$) provides the existence of a solution to equation $F(\Phi,\Psi)=0$ in the domain $S=\{\Phi^2<1, \Psi^2\leq 1\}$. It correspond to Fig.~\ref{fig:fig_1}\,(c). Substitution of $\Phi_*$ into (\ref{eq:line}) yields
\[ \Psi_*=\Phi_*\sqrt{\frac{a_1 h}{a_2 \zeta}}-\frac{u-v}{\sqrt{a_2 \zeta}}=
\frac{w-v}{\sqrt{a_2 \zeta}}\,. \]
If $\Psi_*^2<1$ then straight line (\ref{eq:line}) intersects curve (\ref{eq:curve}) in the square $S$. In this case we have two more real roots of the characteristic polynomial (Fig.~\ref{fig:fig_1}\,(d)).

In summary, we can conclude that the sufficient conditions for the existence of four real roots of $\hat{\chi}(\lambda)=0$ are the inequalities
\[ a_1(\alpha)>\frac{(u-v)^2}{h}, \quad a_2(\alpha)>\frac{(w-v)^2}{\zeta}\,. \]
By definition of $a_1(\alpha)$ and $a_2(\alpha)$, the previous conditions can be rewritten as follows
\begin{equation} \label{eq:hyp-cond}  
\alpha>\alpha_*=\max\Big\{0,\ 3\Big(\frac{(u-v)^2}{\bar{b}h}-\frac{b-\bar{b}}{\bar{b}}\Big),\ 
3\Big(\frac{(w-v)^2}{\bar{b}\zeta}-1\Big) \Big\}\,.
\end{equation}  
For the considered example (see Fig.~\ref{fig:fig_1}) according to (\ref{eq:hyp-cond}) we have $\alpha_*\approx 69.42$. Obviously, equations~(\ref{eq:3L-hyp}) are always hyperbolic for $u=v=w$. The qualitative behaviour of curve (\ref{eq:curve}) in this case corresponds to Fig.~\ref{fig:fig_1}\,(d) and  the curve has an axis of symmetry $\Phi=0$. Straight line (\ref{eq:line}) with the slope $\sqrt{a_1h/(a_2\zeta)}$ passes through the origin and intersects curve (\ref{eq:line}) at four points. In general case, for arbitrary flow parameters, system (\ref{eq:3L-hyp}) is hyperbolic if the parameter $\alpha$ is large enough and condition (\ref{eq:hyp-cond}) is satisfied.

As it was mentioned above, the main advantage of hyperbolic approximation of the dispersive equations is essential simplification of the algorithms of numerical calculation and formulation of the boundary conditions. 

\subsection{Three-layer symmetric flows}\label{subsec:2.3}

Let us consider three-layer stratified flows over a flat bottom ($Z=0$) that possess the property of symmetry with respect to the centre line of the channel $z=H_1=H_0/2$. For such fluid flows $u=w$, $h=\zeta$, $\bar{\rho}=(\rho^-+\rho^+)/2$, and $\bar{b}=b/2$. In this case it is sufficient to consider only the lower part of the flow region ($0\leq y\leq H_1$). Under this assumption, model (\ref{eq:3L-mod}) reduces to a two-layer system and takes the form
\begin{equation} \label{eq:sym-1}   
h_t+(uh)_x=0, \quad (u-v)_t+ \Big(\frac{u^2-v^2}{2}+ \bar{b}h+ \frac{h}{3} 
\frac{d^2 h}{dt^2}+ \frac{1}{6} \Big(\frac{dh}{dt}\Big)^2 \Big)_x=0,
\end{equation}   
where
\[ v=\frac{Q_1-uh}{H_1-h}\,, \quad Q_1=\frac{Q(t)}{2}\,, \quad \frac{d}{dt}= 
\frac{\partial}{\partial t}+u\frac{\partial}{\partial x}\,. \]
As before, we introduce the instantaneous variables $\tilde{h}$ and $\tilde{u}$ such that $d\tilde{h}/dt=\tilde{u}$ and $d\tilde{u}/dt=\alpha \bar{b}(h-\tilde{h})/h$. Then we approximate the dispersive terms in (\ref{eq:sym-1}) according to the first two formulas in (\ref{eq:appr_deriv}). As a result, the system of first-order conservation laws is obtained
\begin{equation} \label{eq:sym-hyp} 
\begin{array}{l} \displaystyle 
h_t+(uh)_x=0, \quad 
(u-v)_t+\Big(\frac{u^2-v^2}{2}+ah- \frac{\alpha}{3}\bar{b}\tilde{h}+ \frac{1}{6}\tilde{u}^2\Big)_x=0, \\[3mm]\displaystyle
(h\tilde{h})_t+(uh\tilde{h})_x=\tilde{u}h, \quad 
(\tilde{u}h)_t+(u\tilde{u}h)_x=\alpha \bar{b}(h-\tilde{h}), \quad a=\Big(1+\frac{\alpha}{3}\Big)\bar{b} \,.
\end{array} 
\end{equation}   
It should be noted that under the assumption of flow symmetry, equations~(\ref{eq:sym-hyp}) also directly follow from system~(\ref{eq:3L-hyp}).

The characteristics of system (\ref{eq:sym-hyp}) can be found explicitly. Let us represent equations~(\ref{eq:sym-hyp}) in form~(\ref{eq:U-form}), where $\bU=(h, s, \tilde{h}, \tilde{u})^{\rm T}$ is the unknown vector (here $s=u-v$), and $\bA(\bU)$ is the matrix of $4\times 4$. Taking into account that the two first equations in (\ref{eq:sym-hyp}) can be rewritten as 
\[ h_t+\Big(u-\frac{sh}{H_1}\Big)h_x+\Big(1-\frac{h}{H_1}\Big) hs_x=0, \quad  s_t+\Big(a-\frac{s^2}{H_1}\Big)h_x+ \Big(u-\frac{sh}{H_1}\Big)s_x-    \frac{\alpha}{3}\bar{b}\tilde{h}_x+   \frac{1}{3}\tilde{u}\tilde{u}_x=0, \]
the eigenvalues of $\bA(\bU)$ are determined by equation
\[ \Big((u-\frac{sh}{H_1}-\lambda\Big)^2- 
\Big(a-\frac{s^2}{H_1}\Big)\Big(1-\frac{h}{H_1}\Big)h\Big)(u-\lambda)^2=0. \]
The roots of this equation (in terms of the variables $u$, $v$ and $h$) are
\[ \lambda_{1,2}=\Big(1-\frac{h}{H_1}\Big)u+\frac{vh}{H_1}\mp 
\sqrt{\Big(a-\frac{(u-v)^2}{H_1}\Big)\Big(1-\frac{h}{H_1}\Big)h}\,, \quad 
\lambda_{3,4}=u.  \]
Therefore, system (\ref{eq:sym-hyp}) is hyperbolic if the inequality 
$a(\alpha)>(u-v)^2/H_1$ is fulfilled. In terms of the parameter $\alpha$, the hyperbolicity condition of equations~(\ref{eq:sym-hyp}) reads
\[ \alpha>\alpha_*=\max \Big\{0,\ 3\,\Big(\frac{(u-v)^2}{\bar{b} H_1}-1\Big) \Big\}\,. \]  
Thus, for system (\ref{eq:sym-hyp}) describing two-layer flows in the Boussinesq approximation taking into account the non-hydrostaticity of one of the layers, the hyperbolicity conditions are formulated in explicit form.

\section{Travelling waves}\label{sec:3}

The solutions to system (\ref{eq:3L-hyp}) in the class of travelling waves are determined from the equations
\begin{equation} \label{eq:tr-wave-hyp} 
\begin{array}{l} \displaystyle 
(u-D)h=J_1={\rm const}, \quad (u-D)\tilde{h}'=\tilde{u}, \quad (u-D)\tilde{u}'=\frac{\alpha\bar{b}}{h}(h-\tilde{h}), \\[3mm]\displaystyle
(w-D)\zeta=J_2={\rm const}, \quad (w-D)\tilde{\zeta}'=\tilde{w}, \quad (w-D)\tilde{w}'=\frac{\alpha\bar{b}}{\zeta}(\zeta-\tilde{\zeta}), \\[3mm]\displaystyle 
(u-D)u'-(v-D)v'+a_1 h'-\frac{\alpha}{3}\,\bar{b}\tilde{h}'+ \frac{1}{3}\tilde{u}\tilde{u}'=0, \\[3mm]\displaystyle 
(w-D)w'-(v-D)v'+a_2 \zeta'-\frac{\alpha}{3}\,\bar{b}\tilde{\zeta}'+ \frac{1}{3}\tilde{w}\tilde{w}'=0\,,
\end{array} 
\end{equation} 
where `prime' denotes the derivative with respect to the variable $\xi=x-Dt$ and $D$ is the constant velocity of the travelling wave. We assume that the bottom is flat ($Z=0$) and the total fluid rate is equal to zero ($Q=0$). 

Let us transform equations (\ref{eq:tr-wave-hyp}) to the normal form. Formulas (\ref{eq:eta_v}) yield
\[ v'=(v-D)\eta^{-1}(h'+\zeta'). \]
Taking this relation into account, the last two equations in (\ref{eq:tr-wave-hyp}) can be written as
\[ \Delta_1 h'-(v-D)^2\eta^{-1}\zeta'=f_1, \quad -(v-D)^2\eta^{-1}h'+\Delta_2 \zeta'=f_2, \]
where
\[ \begin{array}{l} \displaystyle 
\Delta_1=a_1-\frac{(u-D)^2}{h}-\frac{(v-D)^2}{\eta}, \quad
\Delta_2=a_2-\frac{(w-D)^2}{\zeta}-\frac{(v-D)^2}{\eta}, \\[3mm]\displaystyle 
f_1=\frac{C\tilde{u}\tilde{h}}{(u-D)h}, \quad  
f_2=\frac{C\tilde{w}\tilde{\zeta}}{(w-D)\zeta}, \quad C=\frac{\alpha\bar{b}}{3}\,.
\end{array} \]
We solve the previous equations for $h'$ and $\zeta'$. As a result we get
\begin{equation} \label{eq:tr-h-zet} 
\zeta'=\frac{(\Delta_1 h'-f_1)\eta}{(v-D)^2}, \quad 
h'=\frac{\eta^2 f_1\Delta_2+(v-D)^2\eta f_2}{\eta^2\Delta_1\Delta_2-(v-D)^4}.
\end{equation} 
Replacing the last two equations in (\ref{eq:tr-wave-hyp}) with (\ref{eq:tr-h-zet}), we obtain the normal form of this system. 

Further, we are looking for non-trivial solutions of equations~(\ref{eq:tr-wave-hyp}) corresponding to a state of rest as $\xi\to-\infty$:
\begin{equation} \label{eq:const-sol}  
\bU=(h, u, \tilde{h}, \tilde{u}, \zeta, w, \tilde{\zeta}, \tilde{w}) \to 
\bU_0=(h_0, 0, h_0, 0, \zeta_0, 0, \zeta_0, 0). 
\end{equation}   
Here $h_0$ and $\zeta_0$ are constant depths of the non-hydrostatic layers. To construct such a solution, it is necessary first to understand the asymptotic behaviour of the travelling wave solution at negative infinity. Let us consider small perturbations of the constant solution: $\bU=\bU_0+\bU_*$. Substitution this representation into system (\ref{eq:tr-wave-hyp}) and linearisation near solution $\bU_0$ give us the following system for the perturbations
\begin{equation} \label{eq:linear_nonsym}  
\begin{array}{l} \displaystyle 
uh_0-Dh=0, \quad -D\tilde{h}'=\tilde{u}, \quad
-D\tilde{u}'=\alpha\bar{b}(h-\tilde{h})/h_0, \\[3mm]\displaystyle 
w\zeta_0-D\zeta=0, \quad -D^2\eta_0^{-1}h'+\Delta_{20}\zeta'=-CD^{-1}\tilde{w}, \quad
-D\tilde{\zeta}'=\tilde{w}, \\[3mm]\displaystyle
\Delta_{10}h'-D^2\eta_0^{-1}\zeta'=-CD^{-1}\tilde{u}, \quad
-D\tilde{w}'=\alpha\bar{b}(\zeta-\tilde{\zeta})/\zeta_0,
\end{array} 
\end{equation}
where 
\[ \Delta_{10}=a_1-\frac{D^2}{h_0}-\frac{D^2}{\eta_0}, \quad 
\Delta_{20}=a_2-\frac{D^2}{\zeta_0}-\frac{D^2}{\eta_0}, \quad \eta_0=H_0-h_0-\zeta_0, \]
and the `asterisk' symbol is omitted. 

Looking for the solutions of equations (\ref{eq:linear_nonsym}) that vanish at negative infinity in the form
\begin{equation} \label{eq:nu-exp} 
(h, u, \tilde{h}, \tilde{u}, \zeta, w, \tilde{\zeta}, \tilde{w})=
(\hat{h}, \hat{u}, \hat{\tilde{h}}, \tilde{u}, \hat{\zeta}, \hat{w}, \hat{\tilde{\zeta}}, \hat{\tilde{w}})\exp(\nu\xi). 
\end{equation}
Substituting this representation of the solution into equations (\ref{eq:linear_nonsym}) and expressing the unknown amplitudes (denoted with the `hat' symbol) through $\hat{h}$ and $\nu$, we get
\begin{equation} \label{eq:nu-amplitudes} 
\begin{array}{l} \displaystyle 
\hat{u}=\frac{D}{h_0}\hat{h}, \quad \hat{\tilde{h}}=
\frac{\alpha\bar{b}}{\nu^2 D^2 h_0+\alpha\bar{b}}\hat{h}, \quad 
\hat{\tilde{u}}=-\nu D\hat{\tilde{h}}, \quad \hat{w}=\frac{D}{\zeta_0}\hat{\zeta}, \\[3mm]\displaystyle
\hat{\tilde{\zeta}}=\frac{\alpha\bar{b}}{\nu^2 D^2 \zeta_0+\alpha\bar{b}}\hat{\zeta}, \quad \hat{\tilde{w}}=-\nu D\hat{\tilde{\zeta}}, \quad 
\hat{\zeta}=\Big(\Delta_{10}-\frac{\alpha\bar{b}C}{\nu^2D^2h_0+\alpha{\bar{b}}}\Big) \frac{\eta_0}{D^2}\hat{h}.
\end{array} 
\end{equation}
Here $\hat{h}$ is a given amplitude of the small perturbation. A non-trivial ($\hat{h}\neq 0$) solution of system (\ref{eq:linear_nonsym}) in the form (\ref{eq:nu-exp}) exists if the parameter $\nu$ satisfies the equation
\begin{equation} \label{eq:nu-poly} 
\Big(\Delta_{10}-\frac{\alpha\bar{b}C}{\nu^2 D^2 h_0+\alpha\bar{b}}\Big)   
\Big(\Delta_{20}-\frac{\alpha\bar{b}C}{\nu^2 D^2 \zeta_0+\alpha\bar{b}}\Big)=
\frac{D^4}{\eta_0^2}\,. 
\end{equation}  
It is easy to see that (\ref{eq:nu-poly}) reduces to the quadratic equation $\mu^2+A\mu+B=0$ for the variable $\mu=\nu^2$. The coefficients $A$ and $B$ are rather cumbersome and therefore they are not presented here. The existence of a positive root of this equation is a necessary condition for constructing a non-trivial solution to system~(\ref{eq:tr-wave-hyp}) satisfying condition~(\ref{eq:const-sol}) for $\xi\to-\infty$. 
In particular, the fulfilment of this condition is necessary for the construction of solitary waves. We apply asymptotic expressions (\ref{eq:nu-amplitudes}), (\ref{eq:nu-poly}) when the conditions are imposed at $\xi=\xi_0$ in the numerical treatment of system (\ref{eq:tr-wave-hyp}). To solve ODE (\ref{eq:tr-wave-hyp}) numerically, we use the standard ode45 procedure of the MATLAB package.

\begin{figure}[t]
	\begin{center}
		\resizebox{0.6\textwidth}{!}{\includegraphics{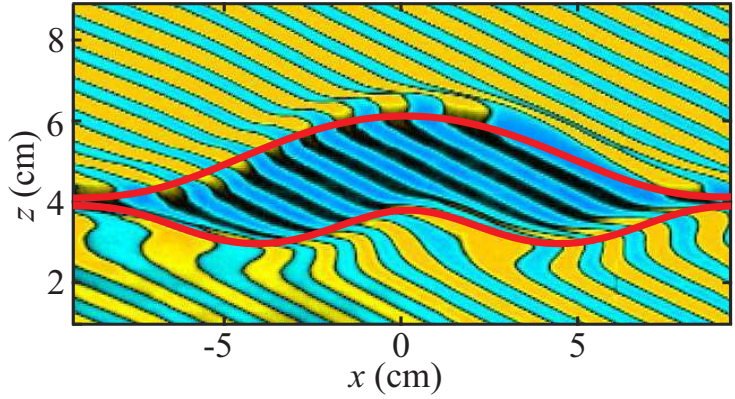}}\\[0pt]
		{\caption{Mode-2 non-symmetric ISW described by system (\ref{eq:tr-wave-hyp}) (bold solid curves correspond to the interfaces $z=h$ and $z=H_0-\zeta$). Coloured picture presents snapshot of experiment \cite{Liapidevskii_2018} (Fig.~6). Blue colour inside of the wave shows the initially coloured fluid trapped by the wave.} \label{fig:fig_2}}
	\end{center}
\end{figure}

An example of a mode-2 non-symmetric internal solitary wave described by equations (\ref{eq:tr-wave-hyp}) is shown in Fig.~\ref{fig:fig_2} by bold solid lines. We choose the following parameters of the flow: $H_0=12$~cm, $\eta_0=0.01\,H_0$, $h_0=H_0/3-\eta_0/2$, $\zeta_0=2H_0/3-\eta_0/2$, $b=5$~cm/s$^2$, $\bar{b}=b/3$, $D=1.61$~cm/s and solve numerically ODE (\ref{eq:tr-wave-hyp}) with conditions at $\xi=\xi_0$ perturbed according to formulas (\ref{eq:nu-amplitudes}) and (\ref{eq:nu-poly}) (we take here $\hat{h}=0.003$, and $\alpha=80$). It should be noted that equation (\ref{eq:nu-poly}) for $\nu^2$ has one positive root. This allow us to uniquely determine the value of $\nu$ and construct the travelling wave solution. The indicated choice of flow parameters corresponds to experiments on the generation of mode-2 non-symmetric internal solitary waves \cite{Liapidevskii_2018,Gavrilov_2013}. In these works, the typical experimental setup with lock-release mechanism was used to generate the internal solitary waves. The interfaces $z=h$ and $z=H_0-\zeta$ obtained according to equations~(\ref{eq:tr-wave-hyp}) are overlaid on a snapshot of a non-symmetric solitary wave of mode-2 (see Fig.~6 in \cite{Liapidevskii_2018}).  Fig.~\ref{fig:fig_2} shows that the constructed solution describes well the non-symmetric solitary wave observed in the experiment. We recall that such a non-symmetric solitary wave exists only for a specific set of governing parameters \cite{Liapidevskii_2018}. In addition, similar non-symmetric solitary waves were recently obtained and studied in \cite{Barros_2020} within the framework of a three-layer long-wave dispersive model. 

\subsection{Symmetric travelling waves}\label{subsec:3.1}

The construction of solutions in the class of travelling waves is simplified in the case of flow symmetry with respect to the centreline $z=H_1=H_0/2$. It follows from system (\ref{eq:sym-hyp}) (or from (\ref{eq:tr-wave-hyp}), (\ref{eq:tr-h-zet}) under assumptions $h=\zeta$, $u=w$, $\tilde{h}=\tilde{\zeta}$, $\tilde{u}=\tilde{w}$, and $\bar{b}=b/2$) that 
\begin{equation} \label{eq:tr-w-sym} 
(u-D)h={\rm const}, \quad \tilde{h}'=\frac{\tilde{u}}{u-D}, \quad 
\tilde{u}'=\frac{\alpha\bar{b}(h-\tilde{h})}{(u-D)h}, \quad 
h'=\frac{f}{\Delta}, 
\end{equation}
where
\[ f=\frac{C\tilde{u}\tilde{h}}{(u-D)h}, \quad 
\Delta=a-\frac{(u-D)^2}{h}-\frac{(v-D)^2}{H_1-h}\,. \]
The study of the asymptotic behaviour of a solution tending to a given state of rest (\ref{eq:const-sol}) at negative infinity is carried out similarly to the previous one and equation (\ref{eq:nu-poly}) takes the form:
\[ \nu^2=\frac{\alpha\bar{b}}{u_0^2h_0} 
\Big(\frac{C}{\Delta_0} -1\Big)\,, \quad
\Big(\Delta_0=C+\bar{b}-\frac{D^2 H_1}{(H_1-h_0)h_0}\,, \quad 
C=\frac{\alpha\bar{b}}{3}\Big)\,. \]
The unknown amplitudes of small perturbations are given by the first three formulas in equations (\ref{eq:nu-amplitudes}). The right-hand side of the equation for $\nu$ is positive, if $\Delta_0>0$ and $C>\Delta_0$. These inequalities give the following condition for the value of wave speed $D$:
\begin{equation} \label{eq:soliton-cond}  
\Big(1-\frac{h_0}{H_1}\Big)\bar{b}h_0<D^2< 
\Big(1-\frac{h_0}{H_1}\Big)\Big(1+\frac{\alpha\varepsilon^2}{3}\Big)\bar{b}h_0 
\end{equation}
Thus, the construction of a solution to equations (\ref{eq:sym-hyp}) in the form of a solitary wave is possible only if condition (\ref{eq:soliton-cond}) is satisfied.

{\bf Remark 1.} The governing equations (\ref{eq:3L-hyp}) (or (\ref{eq:sym-hyp})) are Galilean invariant. Therefore, the study of travelling wave solutions is equivalent to the study of stationary ones.

{\bf Remark 2.} In some cases, it is convenient to rewrite the governing equations (dispersive or hyperbolic) in a dimensionless form by the following scaling
	\[ \begin{array}{l} \displaystyle  
	(x,z)=H_0(x^*,z^*), \quad 
	(h,\eta,\zeta,\tilde{h},\tilde{\zeta})=H_0(h^*,\eta^*,\zeta^*,\tilde{h}^*,\tilde{\zeta}^*), \\[3mm]\displaystyle
	t=t^*H_0/\sqrt{bH_0}, \quad (u,v,w,\tilde{u},\tilde{w})=\sqrt{bH_0}(u^*,v^*,w^*,\tilde{u}^*,\tilde{w}^*).
	\end{array} \]
In these variables the dimensionless buoyancy $b^*=1$ and the total depth $H_0^*=1$.

\section{Numerical results}\label{sec:4}

In this section, we present the results of numerical simulation of the formation and evolution of internal waves. First of all, we show that the results of calculations using the original model~(\ref{eq:3L}) and its modification~(\ref{eq:3L-mod}) practically coincide. This justifies the application of model~(\ref{eq:3L-mod}) to describe the propagation of internal solitary waves. Then we demonstrate that the solution of hyperbolic system~(\ref{eq:3L-mod}) approximates the solution of dispersive equations~(\ref{eq:3L-hyp}). The accuracy of the approximation is determined by the parameter $\alpha$ and spatial resolution. It should be noted that the use of the hyperbolic model allows us to significantly simplify the numerical algorithm and speed up the calculations. Further, we compare the results of modelling the evolution of solitary internal waves of the second mode based on model~(\ref{eq:3L-hyp}) with the calculations performed in \cite{Deepwell_2019} using the fully non-linear Navier--Stokes equations under the Boussinesq approximation. A comparison shows the applicability of model~(\ref{eq:3L-hyp}) for describing the propagation of internal solitary waves and their transformation over an uneven bottom. We also present the results of modelling the interaction of mode-2 symmetric and non-symmetric solitary waves.

\subsection{Numerical methods}\label{sec:4.1}

To solve differential balance laws (\ref{eq:3L-hyp}) numerically, we implement here the Nessyahu--Tadmor second-order central scheme \cite{N_T_1990}
\begin{equation} \label{eq:predictor-corr}
\begin{array}{l}\displaystyle 
\bU_i^{k+1/2}=\bU_i^k-\lambda \mbox{\boldmath$\varphi$}'_i/2+ 
\bF(\bU_i^k)\Delta t/2, \quad\quad (\lambda=\Delta t/\Delta x) \\[2mm]\displaystyle 
\bU_{i+1/2}^{k+1}= (\bU_{i+1}^k+\bU_i^k)/2+ (\bU'_{i+1}-\bU'_i)/8-  \lambda\big(\mbox{\boldmath$\varphi$}(\bU_{i+1}^{k+1/2})- 
\mbox{\boldmath$\varphi$}(\bU_i^{k+1/2})\big)+ \\[2mm]\displaystyle 
\quad\quad\quad +\big(\bF(\bU_{i+1}^k)+\bF(\bU_i^k)\big)\Delta t/2.
\end{array}
\end{equation}
This scheme approximates the systems of balance laws of the form 
\[ \bU_t+(\mbox{\boldmath$\varphi$}(\bU))_x=\bF(\bU). \] 
Here $\bU_i^k =\bU(t^k,x_i)$, $\Delta x$ is the spatial grid spacing, and $\Delta t$ is the time-step satisfying the Courant condition, defined by the velocity of characteristics. The computational domain on the $x$ axis is divided into $N$ cells, the cell centres are denoted by $x_i$. Values $\bU'_i/\Delta x$ and $\mbox{\boldmath$\varphi$}'_i/ \Delta x$ are approximations of the first-order derivatives with respect to $x$. At $t=0$ the initial data $\bU_i^0$ are prescribed. The boundary conditions $\bU_1^k$ and $\bU_N^k$ are also should be specified. Since the values of $\bU_i^k$ are known, one can obtain the conservative variable $\bU_i$ ($i=2,...,N-1$) at the next time step $t^{k+1}$ by formulas (\ref{eq:predictor-corr}).

Since system (\ref{eq:3L-hyp}) consist of eight equations, it is convenient to use central schemes, which do not require exact or approximate solution of the Riemann problem. It should be noted that other schemes (in particular, on non-staggered grids) based on the local Lax--Friedrichs flux are also appropriate here. 

The same method is used to solve dispersive equations~(\ref{eq:3L-num}) for unknown functions $\bU=(h,\zeta,K,R)$, but at each time step we find the velocities $u$ and $w$ in the lower and upper layers from second-order equations~(\ref{eq:K_R}). Let us explain this fact in more details. The numerical approximation of the first order derivatives of any function $\psi$ at $x=x_i$ is given as
\begin{equation} \label{eq:difference-num} 
\bigg(\frac{\partial \psi}{\partial x}\bigg)_i= 
\frac{\psi_{i+1/2}-\psi_{i-1/2}}{\Delta x}, \quad 
\varphi_{i\pm 1/2}=\frac{\psi_{i\pm 1}+\psi_i}{2}\,. 
\end{equation}
Applying finite difference discretization (\ref{eq:difference-num}) one can rewrite ODEs (\ref{eq:K_R}) in the form
\begin{equation} \label{eq:lin-system} 
\alpha_i u_{i-1}- \gamma_i u_i+ \beta_i u_{i+1}- \delta_i w_i= -\phi_i, \quad 
\bar{\alpha}_i w_{i-1}- \bar{\gamma}_i w_i+ \bar{\beta}_i w_{i+1}- \bar{\delta}_i u_i= -\bar{\phi}_i,
\end{equation}
where 
\[ \begin{array}{l} \displaystyle 
\alpha_i=h_{i-1/2}^3, \quad \beta_i=h_{i+1/2}^3, \quad 
\gamma_i=\alpha_i+\beta_i+3\Big(1+\frac{h_i}{\eta_i}\Big)h_i\Delta x^2, \quad
\delta_i= 3\Delta x^2\frac{h_i\zeta_i}{\eta_i} \\[3mm]\displaystyle 
\bar{\alpha}_i=\zeta_{i-1/2}^3, \quad \bar{\beta}_i=\zeta_{i+1/2}^3, \quad 
\bar{\gamma}_i=\bar{\alpha}_i+\bar{\beta}_i+
3\Big(1+\frac{\zeta_i}{\eta_i}\Big)\zeta_i\Delta x^2, \quad
\bar{\delta}_i= \delta_i, \\[3mm]\displaystyle 
\phi_i=3\Big(K_i+\frac{Q}{\eta_i}\Big)h_i\Delta x^2, \quad
\bar{\phi}_i=3\Big(R_i+\frac{Q}{\eta_i}\Big)\zeta_i\Delta x^2.
\end{array} \]
These coefficients correspond to the choice of $n=0$ in equations~(\ref{eq:K_R}) (the case of $n=1$ is similar). All the coefficients are known at time $t^{k+1}$ because the variables $h_i$, $\zeta_i$, $K_i$ and $R_i$ are obtained from the time evolution of conservative variables with the help of scheme~(\ref{eq:predictor-corr}). The only unknowns are the terms $u_i$ and $w_i$ at each node. We introduce the notation
\[ \bu_i= \begin{pmatrix} u_i \\ w_i \end{pmatrix}\,, \quad
\bA_i= \begin{pmatrix} \alpha_i & 0 \\ 0   & \bar{\alpha}_i \end{pmatrix}\,, \quad
\bB_i= \begin{pmatrix} \beta_i & 0 \\ 0   & \bar{\beta}_i \end{pmatrix}\,, \quad
\bC_i= \begin{pmatrix} \gamma_i & \delta_i \\ \bar{\delta}_i & \bar{\gamma}_i \end{pmatrix}\,, \quad 
\bbf_i= \begin{pmatrix} \phi_i \\ \bar{\phi}_i \end{pmatrix}\,. \]
Then equations (\ref{eq:lin-system}) take the form
\begin{equation} \label{eq:three-diag}  
\begin{array}{l} \displaystyle 
\bA_i\bu_{i-1}-\bC_i\bu_i+\bB_i\bu_{i+1}=-\bbf_i, \quad (i=2,...,N-1) \\[2mm]\displaystyle 
\bC_1\bu_1-\bB_1\bu_2=\bbf_1, \quad -\bA_N\bu_{N-1}+\bC_N\bu_N=\bbf_N.
\end{array} 
\end{equation}   
This is a tridiagonal system of linear equations with matrix coefficients that can be solved by a direct (Gauss) method. We apply here a simplified form of Gaussian elimination, which is known as the Thomas algorithm.

\subsection{Comparison of the dispersive and hyperbolic models}\label{sec:4.2}

Let us show that there is almost no difference in calculations on the basis of governing equations (\ref{eq:3L-num}), (\ref{eq:K_R}) with $n=0$ (corresponds to the original model (\ref{eq:3L})) and $n=1$ (proposed modification of equations (\ref{eq:3L})). We also demonstrate that the derived hyperbolic model~(\ref{eq:3L-hyp}) approximates the dispersive equations with high accuracy. 

We perform calculations in the domain $x\in [0,L_0]$ on a uniform grid with $N=1200$ nodes, $L_0=120$~cm. The height of the channel $H_0=12$~cm and the bottom is flat $Z=0$. To start calculations we specify unknown variables $\bU$ in the node points $x=x_i$ at $t=0$ as follows $h=\zeta=0.15\,H_0$ for $x<5$ and $h=4.75$~cm, $\zeta=6.65$~cm for $x>5$. At the initial time $t=0$ the fluid is at rest ($u=v=w=0$). We choose here $b=5$~cm/s$^2$, $\bar{b}=5\,b/12$. At the left and right boundaries we set the impermeability condition. For the hyperbolic model (\ref{eq:3L-hyp}) we take $\alpha=50$ and the initial data for the instantaneous depths and velocities are the same as for $h$, $\zeta$, $u$ and $w$, respectively. 

\begin{figure}[t]
	\begin{center}
		\resizebox{0.98\textwidth}{!}{\includegraphics{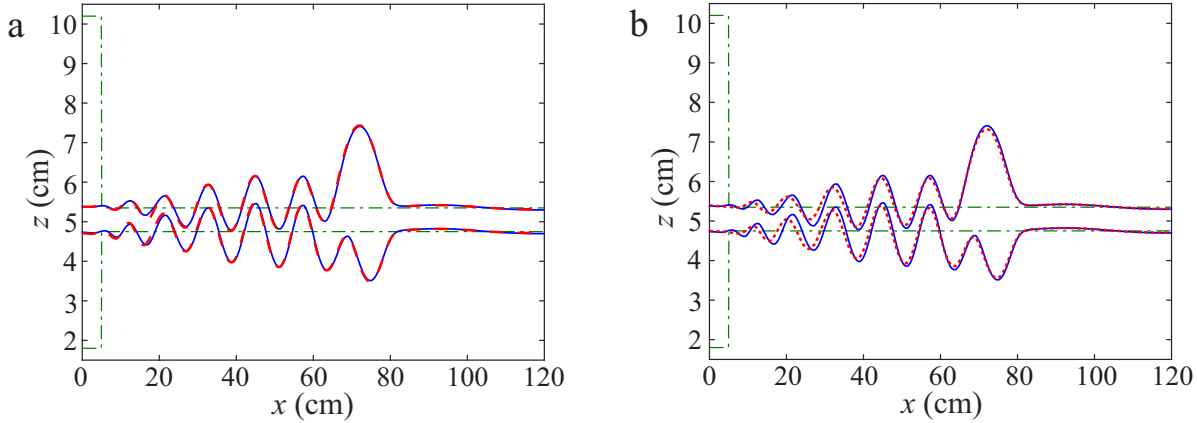}}\\[0pt]
		{\caption{Wave packet formed as a result of evolution of piecewise constant initial data (dash-dotted lines). The interfaces $z=h$ and $z=H_0-\zeta$ are shown at $t=40$~s for a fluid with buoyancy $b=5$~cm/s$^2$ and $\bar{b}=5\,b/12$: (a) solid curves refer to the solution of equations (\ref{eq:3L-mod}), dashed curves correspond to the solution of model (\ref{eq:3L}); (b) solid curves --- system (\ref{eq:3L-mod}), dotted curves --- hyperbolic model (\ref{eq:3L-hyp}).} \label{fig:fig_3}}
	\end{center}
\end{figure}

The packet of internal waves (the interfaces of fluid layers with different density $z=h$ and $z=H_0-\zeta$) generated by the indicated discontinuous initial data is shown in Fig.~\ref{fig:fig_3} at $t=40$~s. A comparison of the solutions of equations (\ref{eq:3L-num}), (\ref{eq:K_R}) for $n=1$ (solid curves) and $n=0$ (dashed) is given in Fig.~\ref{fig:fig_3}\,(a). As we can see, the results of calculating the wave packet for these two dispersive models are practically the same. Minimal differences are noticeable only for the secondary waves, where unsteady effects are manifested. Leading waves propagate at an almost constant speed, that is why they are equally described within these models. Fig.~\ref{fig:fig_3}\,(b) shows that hyperbolic model~(\ref{eq:3L-hyp}) (dashed curves) approximates dispersive equations~(\ref{eq:3L-mod}) (solid lines, coinciding with those shown in Fig.~\ref{fig:fig_3}\,(a)). The accuracy of the approximation depends on the parameter $\alpha$ and spatial resolution. For the minimum values of $\alpha$, which ensure the hyperbolicity of equations~(\ref{eq:3L-hyp}), the leading wave is well approximated. To describe the secondary waves, the value of $\alpha$ should be increased. Moreover, a finer mesh is necessary for large values of the relaxation parameter $\alpha$. A more detailed discussion of the convergence is given in \cite{Chesn_Ng_2019} for a similar problem.

The initial data for the wave packet shown in Fig.~\ref{fig:fig_3} correspond to the experimental conditions described in \cite{Liapidevskii_2018}. In addition, the presented numerical calculations of the generation and propagation of internal waves in a three-layer fluid demonstrate a coincidence with the experimental results (see Fig.~5 in \cite{Liapidevskii_2018}). Thus Fig.~\ref{fig:fig_3} can be considered as verification of the hyperbolic and dispersive models on essentially unsteady flows.

As it is noted in \cite{Favrie_Gavr_2017,Chesn_Ng_2019}, the advantages of hyperbolic approximation include the simplicity of numerical implementation and formulation of non-reflecting conditions at the boundaries. It also allows one to reduce the computation time. In the numerical solution of problems with one non-hydrostatic layer \cite{LeM_G_H_2010,Chesn_Ng_2019}, it is required to solve equations (31) with scalar coefficients at each time step. In our case, these are $2\times 2$ matrices, which significantly increases the computation time. In particular, for the presented example (see Fig.~\ref{fig:fig_3}), the calculation time by the dispersive model (\ref{eq:3L-mod}) is $T_c=540.95$ (in seconds). The number of iterations (time steps) $M=11762$ is needed to reach the final time $t=40$~s. For the original dispersive model~(\ref{eq:3L}), the results are comparable ($T_c=607.27$~s). The calculation time by the hyperbolic model~(\ref{eq:3L-hyp}) with $\alpha=50$ is $T_c=64.48 $~s (it takes $M=20507$ iterations). All calculations are carried out on the same computer on a uniform grid with $N=1200$ nodes. Obviously, the use of hyperbolic approximation allows one to reduce the computation time.

\subsection{Transformation of solitary waves over a broad ridge}\label{sec:4.3}

In \cite{Deepwell_2019} the passage of a mode-2 ISW over a broad isolated ridge was explored using both numerical simulations and laboratory experiments. We compare the results of modelling the propagation and transformation of ISWs based on the hyperbolic model (\ref{eq:3L-hyp}) with the numerical results obtained in \cite{Deepwell_2019} using the fully non-linear two dimensional Navier--Stokes equations under the Boussinesq approximation. 

The initial data generating internal waves in a three-layer fluid are similar to the previous test. Following \cite{Deepwell_2019} (case 3010) we take $H_0=0.3$~m and the form of the bottom is 
\[ Z(x)=A\exp(-(x-x_0)^2/\sigma^2), \] 
where $A=H_0/3$, $\sigma=0.6$~m and $x_0=8\,\sigma$. At $t=0$ we choose $h=\zeta=H_0/3$ for $x<H_0$ and $\zeta=7\,H_0/15$, $h=\zeta-Z(x)$ for $x>H_0$. At the initial time $t=0$ the fluid is at rest. The fluid stratification is defined as follows $\rho^\pm=\bar{\rho}\mp\Delta\rho/2$. In \cite{Deepwell_2019} the ratio $\rho_*=\Delta\rho/\bar{\rho}$ is slightly different in experiment ($\rho_*=0.0021$) and numerical simulation ($\rho_*=0.0023$). In our calculation we take $\rho_*=0.0019$ (it corresponds to the buoyancy $b=0.188$~m/s$^2$ and $\bar{b}=b/2$). Such a choice gives a leading wave velocity close to that was obtained in the numerical simulation \cite{Deepwell_2019}. We perform calculations in the domain $x\in [0,L_0]$, $L_0=10$~m on a uniform grid with $N=4000$ nodes. The relaxation parameter is $\alpha=25$ (increasing $\alpha$ and improving spatial resolution have practically no effect on the results). We use here the impermeability condition at the boundaries of the computational domain.

\begin{figure}[t]
	\begin{center}
		\resizebox{.85\textwidth}{!}{\includegraphics{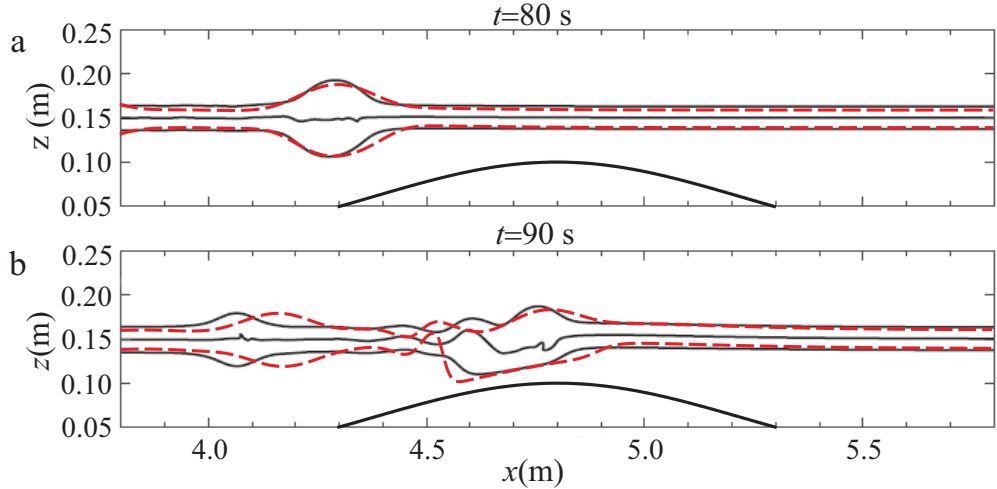}}\\[0pt]
		{\caption{The boundaries of the fluid layers $z=h+Z$ and $z=H_0-\zeta$ obtained by model (\ref{eq:3L-hyp}) at $t=80$~s and $t=90$~s (dashed curves). Solid curves --- time series of isopycnals for internal waves passing over broad topography (Fig.~11 (a, b) from \cite{Deepwell_2019}). } \label{fig:fig_4}} 
	\end{center}
\end{figure}

Fig.~\ref{fig:fig_4} shows the evolution of the incident mode-2 internal solitary wave over broad ridge (bold solid curves). The fluid layer boundaries $z=h$ and $z=H_0-\zeta$ at time instants $t=80$ and $t=90$ seconds obtained on the basis of model (\ref{eq:3L-hyp}) are shown by the dashed curves. These curves are overlaid on Fig.~11\,(a,\,b) from \cite{Deepwell_2019}, which present the time series of isopycnals (solid curves) obtained as a result of solving the two-dimensional Navier--Stokes equations. It is not surprising that we do not observe complete coincidence of the calculation results, since significantly different models are used. In addition, we can not give a comparison for the middle isopycnal, because we use three-layer equations. We also note that model~(\ref{eq:3L-hyp}) does not include any dissipative terms. Nevertheless, the simpler one-dimensional model (\ref{eq:3L-hyp}) gives results similar to the two-dimensional Navier--Stokes equations. In particular, the incident wave is stable when it reaches the ridge where the lower portion of the wave becomes obstructed, which causes it to lag behind the upper portion. As in \cite{Deepwell_2019}, the further evolution of the flow leads to the formation of a series of trailing mode-1 waves and the transmitted wave amplitude becomes noticeably smaller than the incident wave.

\subsection{Interaction of mode-2 internal solitary waves}\label{sec:4.4}

In the framework of the hyperbolic model (\ref{eq:3L-hyp}), we perform numerical modelling of the interaction of ISWs moving towards each other. In this section, we use dimensionless variables and consider flows above a flat bottom ($Z=0$). Without loss of generality, we assume $H_0=1$, $Q=0$, $b=1$. We choose here $\alpha=50$. All subsequent calculations are carried out on a uniform grid with the number of nodes $N=3000$. At the boundaries of the computational domain impermeability conditions are set.

\begin{figure}[t]
	\begin{center}
		\resizebox{1\textwidth}{!}{\includegraphics{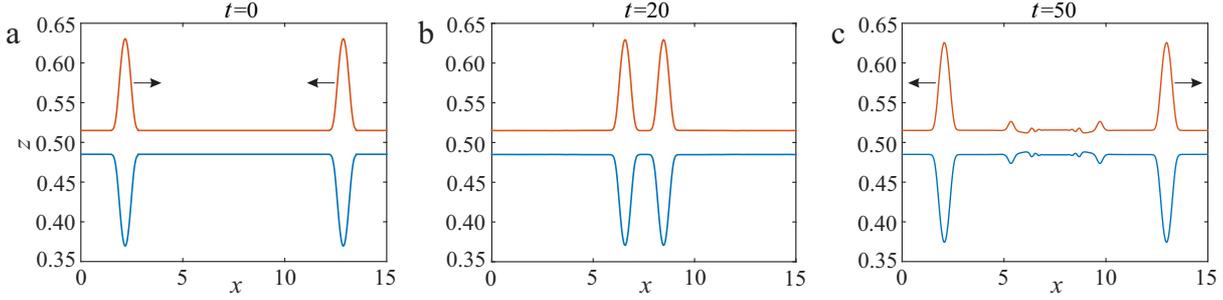}}\\[0pt]
		{\caption{Evolution and interaction of mode-2 two internal solitary waves. The interfaces $z=h$ and $z=H_0-\zeta$ are shown at $t=0$ (a), $t=20$ (b), and $t=50$ (c).} \label{fig:fig_5}} 
	\end{center}
\end{figure}

Firstly, we consider the interaction of mode-2 ISWs in a three-layer fluid. To do this, we construct a non-trivial solution of equations (\ref{eq:tr-wave-hyp}) satisfying condition (\ref{eq:const-sol}). As it is noted above, if there is symmetry about the centre line of the channel, the travelling wave is determined by solving equations (\ref{eq:tr-w-sym}). In this case, it is necessary to fulfil condition (\ref{eq:soliton-cond}) for the wave propagation velocity. We take here $\bar{b}=0.5$, $\eta_0=0.03$, $h_0=\zeta_0=(1-\eta_0)/2$, $\hat{h}=-0.0003$ and $D=0.22$. With the indicated parameters, ODEs~(\ref{eq:tr-w-sym}) have a solution in the form of a soliton having an amplitude $A_s=0.1154$ and a length $L=1.30$. In the computational domain $x\in [0,15]$ at $t=0$ we take two constructed solitons as the initial data (Fig.~\ref{fig:fig_5}\,(a)). For a soliton located closer to the right boundary, the wave velocity $D$ is replaced by $-D$. The evolution of these initial data, calculated on the basis of non-stationary governing equations~(\ref{eq:3L-hyp}), is shown in Fig.~\ref{fig:fig_5}\,(b,\,c). Before the collision, ISWs move towards each other with the preservation of shape. After the interaction, the waves take their original shape, but several small-amplitude waves form between them. It should be noted that small wave perturbations after the ISWs interaction are also observed in calculations based on the two-dimensional Navier--Stokes equations \cite{Maderich_2017}.

\begin{figure}[t]
	\begin{center}
		\resizebox{.85\textwidth}{!}{\includegraphics{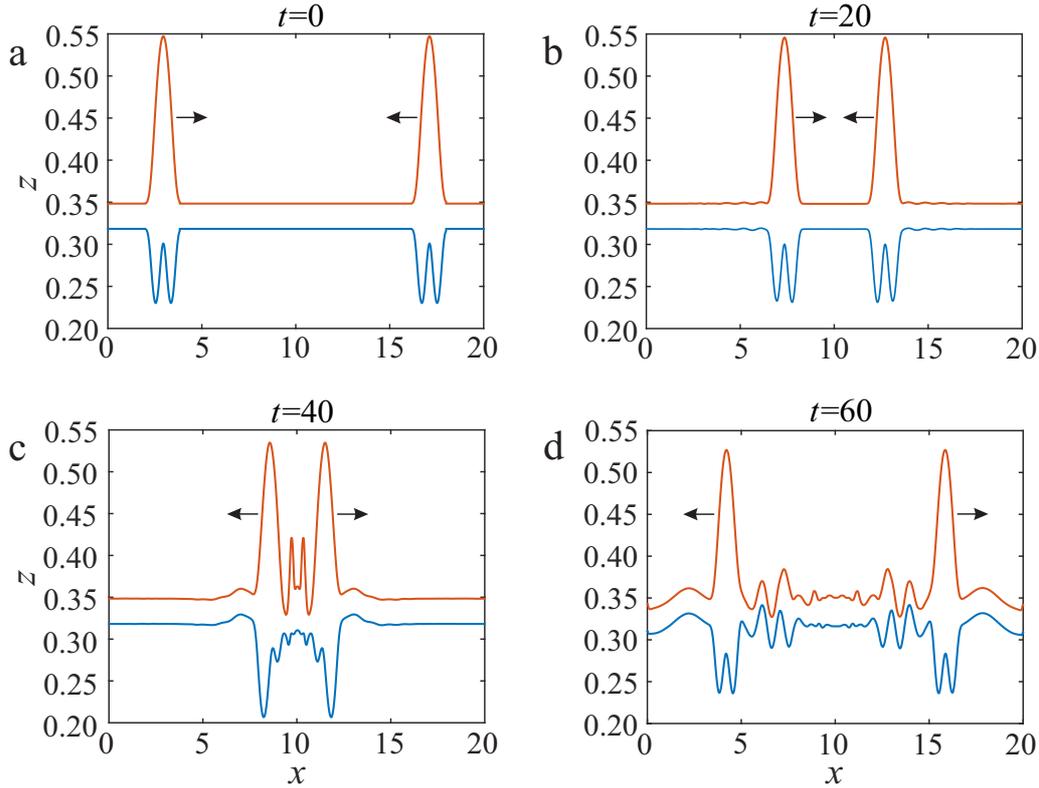}}\\[0pt]
		{\caption{Time series of non-symmetric mode-2 internal solitary waves evolution: $t=0$ (a), $t=20$ (b), $t=40$ (c), and $t=60$ (d).} \label{fig:fig_6}} 
	\end{center}
\end{figure}

Let us consider the interaction of two non-symmetric solitary waves obtained above (see Fig.~\ref{fig:fig_2}). We choose for this test $\bar{b}=1/3$, $\eta_0=0.03$, $h_0=1/3-\eta_0/2$, $\zeta_0=2/3-\eta_0/2$. The remaining parameters are the same as in the previous example. In this case, as a result of solving equations~(\ref{eq:tr-wave-hyp}) with conditions~(\ref{eq:const-sol}), we obtain an non-symmetric solitary wave of length $L=1.83$ and amplitude $A=0.1988$ (for the upper interface $z=H_0-\zeta$). At the initial time, in the vicinity of the left and right boundaries of the computational domain $x\in [0,20]$, we place two such waves moving towards each other at a speed of $D=0.22$. Fig.~\ref{fig:fig_6} shows the constructed initial data and the corresponding numerical solution of equations~(\ref{eq:3L-hyp}) at different time moments. Before the collision, non-symmetric solitary waves move with the preservation of shape, but behind them trains of small-amplitude waves are formed (Fig.~\ref{fig:fig_6}\,(b)). After the interaction, non-symmetric internal solitary waves take the form close to the original one. However, behind the leading waves and in front of them, a series of trailing mode-1 waves are formed. The amplitude of these perturbations is noticeably smaller than the amplitude of the leading waves, but significantly larger than in the previous example when considering the interaction of symmetrical solitary waves. 

\section{Conclusions}\label{sec:5}

We consider three-layer shallow water models (\ref{eq:3L}) and (\ref{eq:3L-mod}) describing dynamics of large-amplitude internal waves in a stratified fluid. The main feature of these models is that non-hydrostatic effects are taken into account only in two outer layers (the intermediate layer is hydrostatic). Such models allow one to study soliton-like solutions representing ISWs of the first and second modes. Using additional `instantaneous' variables and a relaxation parameter \cite{LT00,Liapidevskii_2008}, we obtain first-order hyperbolic system (\ref{eq:3L-hyp}) that approximates three-layer dispersive model (\ref{eq:3L-mod}). We derive an equation determining the velocities of the characteristics of this system and, applying a geometric interpretation (see Fig.~\ref{fig:fig_1}), we prove that all of them are real for a sufficiently large relaxation parameter. The advantage of the hyperbolic model is simpler numerical implementation and formulation of boundary conditions. In contrast to the solution of dispersive equations, in this case there is no need for a time-consuming operation for solving a system of second order ODEs at each time step. We study the solutions of the proposed hyperbolic model in the form of travelling waves (\ref{eq:tr-wave-hyp}) and formulate the necessary conditions (\ref{eq:nu-poly}) for the existence of a solitary wave. We present a solution of the hyperbolic model describing a mode-2 non-symmetric ISW (Fig.~\ref{fig:fig_2}). This solution is verified by comparison with laboratory experiment \cite{Gavrilov_2013,Liapidevskii_2018}. In the case of symmetry with respect to the centre line of the channel, the construction of a travelling wave solution is essentially simplified.

Further, we apply the dispersive and hyperbolic systems to describe time-dependent flows of a three-layer stratified fluid. Firstly, we describe the main features of the numerical implementation of the considered models. To solve both the hyperbolic and dispersive systems, we apply the Nessyahu--Tadmor second order central scheme~\cite{N_T_1990}. However, for the dispersive system at each time step  it is necessary to solve tridiagonal system (\ref{eq:three-diag}) with matrix coefficients. Obviously, this significantly slows down the calculations. Therefore, the use of hyperbolic equations is preferable. Then we compare the results of numerical modelling based on the three-layer Green--Naghdi equations, their modification and hyperbolic approximation. The comparison made shows that all these models give almost the same result (see Fig.~\ref{fig:fig_3}). The performed simulation of the propagation of mode-2 ISWs and their transformation over a broad isolated ridge in the framework of 1D hyperbolic model is in a good agreement with the results obtained in \cite{Deepwell_2019} on the basis of 2D Navier--Stokes equations (Fig.~\ref{fig:fig_4}). Thus, the depth-averaged three-layer model makes it possible to correctly describe the main features of the solitary waves decaying over an uneven bottom. We also present the numerical results of modelling the interaction of two mode-2 symmetric and non-symmetric ISWs. The behaviour of symmetrical solitary waves is fairly common (Fig.~\ref{fig:fig_5}), while in the case of non-symmetric waves interaction, a series of trailing waves of mode-1 are formed (Fig.~\ref{fig:fig_6}). Further development of the layered flow model consists in taking into account the mixing process between the fluid layers and the formation of vortices during the breaking of the internal waves.

\section*{Acknowledgements}
This work was supported by the Russian Science Foundation (project 20-11-20189).


\begin{thebibliography}{}

\bibitem{Helfrich_2006}
K.R. Helfrich, W.K. Melville, Annu. Rev. Fluid Mech. \textbf{38}, 395 (2006)

\bibitem{Jackson_2004}
C.R. Jackson, An Atlas of Internal Solitary-like Waves and their Properties, 2nd ed. (Global Ocean Associates, 2004), see www.internalwaveatlas.com

\bibitem{Shroyer_2010} 
E.L. Shroyer, J.N. Moum, J.D. Nash, J. Geophys. Res. \textbf{115}, C07001 (2010)

\bibitem{Ramp_2012}
S.R. Ramp, Y.J. Yang, D.B. Reeder, F.L. Bahr, J. Geophys. Res. \textbf{117}, C03043 (2012)

\bibitem{Silva_2015} 
J.C.B. da Silva, M.C. Buijsman, J.M. Magalhaes, Deep-Sea Research I \textbf{99}, 87 (2015)

\bibitem{Akylas_1992}
T.R. Akylas, R.H.J. Grimshaw, J. Fluid Mech. \textbf{242}, 279 (1992)

\bibitem{Choi_2000}
W. Choi, Modelling of strongly non-linear internal gravity waves, in Proc. of the 4th Intern. Conf. on Hydrodynamics, Yokohama, 7--9 Sept. 2000 (Yokohama, 2000).

\bibitem{Antuono_2009}
M. Antuono, V. Liapidevskii, M. Brocchini, Stud. Appl. Math. \textbf{122}, 1 (2009)

\bibitem{Salloum_2012} 
M. Salloum, O.M. Knio, A. Brandt, Phys. Fluids \textbf{24}, 016602 (2012)

\bibitem{Zhang_2018} 
P. Zhang, Z. Xu, Q. Li, B. Yin, Y. Hou, A. K. Liu, Nonlin. Processes Geophys. \textbf{25}, 441 (2018)

\bibitem{Brandt_2014}
A. Brandt, K.R. Shipley, Phys. Fluids \textbf{26}, 046601 (2014)

\bibitem{Carr_2015}
M. Carr, P.A. Davies, R.P. Hoebers, Phys. Fluids \textbf{27}, 046602 (2015)

\bibitem{Yuan_2018} 
C. Yuan, R. Grimshaw, E. Johnson, J. Fluid Mech. \textbf{836}, 238 (2018)

\bibitem{Maderich_2017} 
V. Maderich, K.T. Jung, K. Terletska, K.O. Kim, Nonlin. Processes Geophys. \textbf{24}, 751 (2017)

\bibitem{Deepwell_2017}
D. Deepwell, M. Stastna, M. Carr, P.A. Davies, Phys. Fluids \textbf{29}, 076601 (2017)

\bibitem{Deepwell_2019}
D. Deepwell, M. Stastna, M. Carr, P.A. Davies, Phis. Rev. Fluids \textbf{4}, 094802 (2019)

\bibitem{Terletska_2016} 
K. Terletska, K.T. Jung, T. Talipova, V. Maderich, I. Brovchenko, R. Grimshaw, Phys. Fluids \textbf{28}, 116602 (2016)

\bibitem{Gavrilov_2012}
N. Gavrilov, V. Lyapidevskii, K. Gavrilova, Nonlin. Processes Geophys. \textbf{19}, 265 (2012)

\bibitem{Liapidevskii_2017}
V.Yu. Liapidevskii, V.V. Novotryasov, F.F. Khrapchenkov, I.O. Yaroshchuk, J. Appl. Mech. Tech. Phys. \textbf{58}, 809 (2017)

\bibitem{Liapidevskii_2018}
V.Yu. Liapidevskii, N.V. Gavrilov, in The Ocean in Motion, ed. by M.G. Velarde et al. (Springer Oceanography, 2018), pp.\,87--108

\bibitem{Kukarin_2019}
V.F. Kukarin, V.Yu. Liapidevskii, F.F. Khrapchenkov, I.O. Yaroshchuk, Fluid Dynamics \textbf{54}, 329 (2019)

\bibitem{Barros_2020} 
R. Barros, W. Choi, P.A. Milewski, J. Fluid Mech. \textbf{883}, A16 (2020)
`
\bibitem{LeM_G_H_2010}
O. Le Metayer, S. Gavrilyuk, S. Hank, J. Comput. Phys. \textbf{229}, 2034 (2010)

\bibitem{LT00} 
V.Yu. Liapidevskii, V.M. Teshukov, Mathematical Models of Propagation of Long Waves in a Non-Homogeneous Fluid, (Siberian Branch of Russian Academy of Sciences, Novosibirsk, 2000) [in Russian] 

\bibitem{Liapidevskii_2008} 
V.Yu. Liapidevskii, K.N. Gavrilova, J. Appl. Mech. Tech. Phys. \textbf{49}, 34 (2008)   

\bibitem{Favrie_Gavr_2017}
N. Favrie, S. Gavrilyuk, Nonlinearity \textbf{30}, 2718 (2017)

\bibitem{Chesn_Ng_2019}
A.A. Chesnokov, T.H. Nguyen, Comput. Fluids \textbf{189}, 13 (2019).

\bibitem{Serre_1953} F. Serre, Contribution \`{a} l'\'{e}tude des \'{e}coulements permanents et variables dasn les cannaux, (Houille Blanche 8, 1953), pp.\,374--388.

\bibitem{G_L_Ch_2019}
S.L. Gavrilyuk, V.Yu. Liapidevskii, A.A. Chesnokov, Europ. J. Mech. B/Fluids \textbf{73}, 157 (2019) 

\bibitem{Gavrilov_2013} 
N.V. Gavrilov, V.Yu. Liapidevskii, Z.A. Liapidevskaya, Fundam. i Prikl. Gidrofizika \textbf{6}, 25 (2013) [in Russian]

\bibitem{Ovs_79}
L.V. Ovsyannikov, J. Appl.Mech. Tech. Phys. \textbf{20}, 127 (1979)  

\bibitem{Chesn_2017}
A.A. Chesnokov, G.A. El, S.L. Gavrilyuk, M.V. Pavlov, SIAM J. Appl. Math. \textbf{77}, 1068 (2017) 

\bibitem{N_T_1990} 
H. Nessyahu, E. Tadmor, J. Comp. Phys. \textbf{87}, 408 (1990)

\end{thebibliography}
\end{document}